%% file: Sec_Main.tex
\documentclass[runningheads]{llncs}
\input{setting2_packages}

\input{setting1_variable}

\input{setting3_FIXED}
\begin{document}
\title{Detecting and Characterizing Low and No Functionality Packages 
in the NPM Ecosystem}

\titlerunning{Detecting Low and No Functionality}
%
\author{
Napasorn Tevarut\inst{1} \and
Brittany Reid \inst{2} \Envelope  \and
Yutaro Kashiwa\inst{2} \and
Pattara Leelaprute\inst{1} \and
Arnon Rungsawang\inst{1} \and
Bundit Manaskasemsak\inst{1} \and
Hajimu Iida\inst{2}
}
%
\authorrunning{N. Tevarut et al.}
%
\institute{Faculty of Engineering, Kasetsart University, Thailand
\and Nara Institute of Science and Technology (NAIST), Japan\\
\email{Corresponding Author\Envelope{}: brittany.reid@naist.ac.jp}
}
\maketitle              
\begin{abstract}

Trivial packages, small modules with low functionality, are common in the npm ecosystem and can pose security risks despite their simplicity. This paper refines existing definitions and introduce data-only packages that contain no executable logic. A rule-based static analysis method is developed to detect trivial and data-only packages and evaluate their prevalence and associated risks in the 2025 npm ecosystem. The analysis shows that 17.92\% of packages are trivial, with vulnerability levels comparable to non-trivial ones, and data-only packages, though rare, also contain risks. The proposed detection tool achieves 94\% accuracy (macro-F1 0.87), enabling effective large-scale analysis to reduce security exposure. This findings suggest that trivial and data-only packages warrant greater attention in dependency management to reduce potential technical debt and security exposure.

\keywords{Software Libraries  \and npm Ecosystem \and security vulnerability.}
\end{abstract}
\input{Sec1}

\input{Sec2}
\input{Sec3}

\input{Sec4}
\input{Sec5}

\input{Sec6}

\subsection*{\ackname}
We gratefully acknowledge the financial support of JSPS KAKENHI grants (JP24K02921, JP25K21359), as well as JST PRESTO grant (JPMJPR22P3), ASPIRE grant (JPMJAP2415), and AIP Accelerated Program (JPMJCR25U7).

\bibliographystyle{splncs04}
\bibliography{references}

\end{document}

%% file: setting2_packages.tex
\usepackage[T1]{fontenc}
\usepackage{amsmath,amsfonts}
\usepackage{booktabs}
\usepackage{fancyvrb}
\usepackage{xurl}
\usepackage{url}
\usepackage{algorithmic}
\usepackage{textcomp}
\usepackage{xspace}
\usepackage[aboveskip=2pt]{subcaption}
\usepackage{tabularx}
\usepackage{balance}
\usepackage{framed}
\usepackage{chngpage}
\usepackage[many]{tcolorbox}
\usepackage{pgfplots}
\usepackage{siunitx}
\usepackage{enumitem}
\usepackage{multirow}
\usepackage{listings}
\usepackage{bbding}

\usepackage{hyperref}
\usepackage[table,xcdraw]{xcolor}

\usepackage{graphicx}
\usepackage{subcaption}

%% file: setting1_variable.tex
\newcommand{\openAPRs}{54}
\newcommand{\mergedAPRs}{475}
\newcommand{\closedAPRs}{92}

\pgfmathsetmacro{\collectedAPRs}{int(\closedAPRs+\openAPRs+\mergedAPRs)}
\pgfmathsetmacro{\studiedAPRs}{int(\mergedAPRs+\closedAPRs)}

\definecolor{codegray}{gray}{0.95}

\lstdefinestyle{mypackagecode}{
  backgroundcolor=\color{codegray},
  basicstyle=\ttfamily\scriptsize,
  breaklines=true,
  frame=single,
  columns=fullflexible,
  aboveskip=-0.5em,
  belowskip=-1.5em,
}

\pgfmathsetmacro{\calcMergeRateAPR}{(\mergedAPRs/\studiedAPRs)*100}

\pgfmathsetmacro{\calcRejectRateAPR}{1-\calcMergeRateAPR}

\newcommand{\directMergedAPRs}{208}
\pgfmathsetmacro{\calcDirectMergeRateAPR}{(\directMergedAPRs/\mergedAPRs)*100}
\pgfmathsetmacro{\calcInDirectMergeRateAPR}{(1-(\directMergedAPRs/\mergedAPRs))*100}

\newcommand{\mergedHPRs}{521}
\newcommand{\closedHPRs}{46}

\pgfmathsetmacro{\collectedHPRs}{int(\closedHPRs+\mergedHPRs)}
\pgfmathsetmacro{\studiedHPRs}{int(\mergedHPRs+\closedHPRs)}

\pgfmathsetmacro{\calcMergeRateHPR}{(\mergedHPRs/\studiedHPRs)*100}
\pgfmathsetmacro{\calcRejectRateHPR}{1-\calcMergeRateHPR}

\newcommand{\directMergedHPRs}{219}
\pgfmathsetmacro{\calcDirectMergeRateHPRs}{(\directMergedHPRs/\mergedHPRs)*100}

%% file: setting3_FIXED.tex
\definecolor{darkgreen}{rgb}{0, 0.5, 0} 
\definecolor{whitesmoke}{rgb}{0.99, 0.99, 0.99} 

\def\Underline{\setbox0\hbox\bgroup\let\\\endUnderline}
\def\endUnderline{\vphantom{y}\egroup\smash{\underline{\box0}}\\}
\def\|{\verb|}

\newcounter{findings_no}

\usepackage{listings}
\definecolor{backcolour}{rgb}{0.95,0.95,0.92}
\lstdefinelanguage{diff}{
  morecomment=**[f][\color{red}]{-},         
  morecomment=**[f][\color{darkgreen}]{+},       
  moredelim=**[is][\bfseries]{@@}{@@},
}
\definecolor{backcolour}{rgb}{0.95,0.95,0.92}
\lstdefinelanguage{commit}{ 
  breakindent = 0pt,
  numbers=none,
  backgroundcolor=\color{white},
  frame=single,
  xleftmargin=3.5em,
  numbersep=0em,
  xrightmargin=1.5em,
}

\usepackage[many]{tcolorbox}  
\tcbuselibrary{listings,breakable}
\definecolor{main}{HTML}{D0D3D4}    
\definecolor{sub}{HTML}{D0D3D4}     
\newtcolorbox{dbox}{
    left=1pt,right=1pt,top=1pt,bottom=1pt,
    enhanced, 
    boxrule = 0pt,
    enlarge top by=5pt,
    enlarge bottom by=3pt,
  }
\tcbset{
    left=3pt,right=3pt,top=3pt,bottom=3pt,
    sharp corners,
    before skip = 0.1pt,    
    after skip = 0.1pt      
}



%% file: Sec1.tex
\section{Introduction}
JavaScript is one of the most widely used programming languages. It's associated package manager, which enables the download, installation and updating of third-party dependencies, NPM (Node Package Manager), host over 3.5 million packages as of July 2025 and continues to grow rapidly\footnote{\url{https://replicate.npmjs.com/}}. While library use accelerates development, it also introduces risks by creating longer dependency chains that increase indirect vulnerabilities and maintenance overhead. Node.js developers often rely on small, low-functionality (‘trivial’) packages. 
Despite their apparent simplicity, such packages can create deep dependency chains \cite{Abdalkareem,micro}, increasing vulnerability and maintenance risks \cite{micro}, as seen in the \texttt{left-pad} incident \cite{Abdalkareem,micro,untrivaility} that disrupted major platforms.

Prior research has defined trivial, low functionality packages based on lines of code and cyclomatic complexity \cite{Abdalkareem}, or function count \cite{micro}, and shown they can be as risky as larger ones due to transitive dependencies \cite{untrivaility,VulandDep}. However, existing work has not investigated the security of trivial packages themselves. Additionally, The study identified a set of packages that contain no functionality -- data-only packages. These contain no executable logic and serve solely as containers for static values or datasets, such as \texttt{color-names}, which maps color names to hex codes and \texttt{const-log10e}, which exports a single numeric constant. While seemingly harmless, they can still introduce risks through bundled metadata, configuration content, or dependency chains—similar to trivial packages. However, no systematic study has addressed data-only packages in the npm ecosystem or proposed automated detection methods for them.

This gap matters since many developers are unaware that trivial or data-only packages increase security risks~\cite{Abdalkareem}. In some cases, developers may include these packages in their dependency chain without realizing it. To address this issue, The study propose a rule-based detection method and conduct an empirical study on their prevalence and risks in the npm ecosystem in 2025. The result indicate that 17.92\% of npm packages are trivial and remain widely uses today, with vulnerability levels similar to non-trivial ones. This paper also define data-only packages as those with no functions, complexity per file ~$\leq$~ 1, and no import statements; although logic-free, they can still carry vulnerabilities. Using both definitions, our detection tool achieved 94\% accuracy, enabling effective large-scale analysis.

%% file: Sec2.tex
\section{Related Work}
In this section, we discuss studies related to this work, primarily on trivial packages in the npm ecosystem.
Abdalkareem et al. \cite{Abdalkareem} defined trivial packages as those with $\leq$35 lines of code and cyclomatic complexity $\leq$10, while Kula et al. \cite{micro} focused on micro-packages, identifying them by function count $\leq$1. Both studies emphasize that, despite their minimal functionality, such packages can create deep dependency chains.
Chowdhury et al. \cite{untrivaility} extended this perspective by showing that trivial packages can have non-trivial impacts on security and stability in large-scale JavaScript projects.
Similarly, Decan et al. \cite{VulandDep} examined how vulnerabilities propagate through npm’s dependency network, revealing that security issues can affect both small  and large packages via transitive dependencies. These findings highlight that trivial packages can present significant risks despite their minimal features.

While these studies address the prevalence and risks of small or trivial packages, none explicitly characterize data-only packages—packages that contain no executable logic. 
Our work extends this research by defining data-only packages, assessing their prevalence and risks, and proposing an automated detection method.

%% file: Sec3.tex
\section{Dataset}
\label{section:dataset}

To support our study on low functionality and no functionality packages in the NPM ecosystem, we constructed a dataset by randomly sampling packages from the public NPM registry.

\begin{table}[h]
    \centering
    \caption{Overview of the dataset.}
    \begin{tabular}{lr}
        \toprule
        \textbf{Subset} & \textbf{Number of Packages} \\
        \midrule
        NPM Packages as of July 2025 & 3.5 million \\
        Random Sample & 3281 \\
        After Filtering & 3220 \\
        \\
        Low Functionality (Trivial) &  577\\
        No Functionality (Data-only) & 40 \\
        \\ 
        Vulnerability Count & 5660 \\
        \bottomrule
    \end{tabular}
    \label{tab:my_label}
\end{table}

A total of 3,281 packages were randomly selected using data from \textbf{npms.io}.\footnote{\url{https://npms.io/}}

The selection was designed to encompass a wide range of functionality, sizes, and popularity levels. The dataset includes both newly published and long-standing packages, reflecting the current state of the ecosystem as of July 2025. 

For each package, the name, GitHub repository link, download count, and score were obtained from the npms.io API. In addition 
the repository source files were retrieved using the \texttt{npm install} command,and the dataset excludes irrelevant files such as tests, type declarations, and distribution/build directories to ensure consistency.

Furthermore, 61 packages that contained no measurable source code were filtered out, as they did not provide any implementation logic and would not contribute meaningfully to the analysis.

The dataset also contains five calculated package metrics for the remaining 3,220 packages; 1) lines of code (LOC), 2) cyclomatic complexity, 3) function count, 4) dependency count, and 5) known vulnerabilities. First, LOC was calculated using \texttt{cloc},\footnote{\url{https://www.npmjs.com/package/cloc}} excluding comments and empty lines. Cyclomatic complexity and function count were calculated using \texttt{typhonjs-escomplex}~\cite{Visuallize}. The dependency count was calculated as the total number of transitive dependencies (both direct and indirect) by recursively traversing the dependency graph of each package \cite{Abdalkareem}. Finally, known vulnerabilities were collected using the \texttt{npm audit} command in JSON output mode, capturing the number and severity of known vulnerabilities for each package.

%% file: Sec4.tex
\section{Results}

\subsection{RQ1: How prevalent are trivial packages in the ecosystem?}

The analysis examined the prevalence, popularity, and download counts of trivial libraries within the sample of 3,220 mined NPM packages.
To effectively explore and detect trivial packages, 
a clear definition of what constituted a trivial package was established. Following the definition from Abdalkareem et al. \cite{Abdalkareem}, classifying packages as trivial if they had LOC $\leq$ 35 and complexity $\leq$ 10. Among the 3,220 mined NPM packages, 577 packages (17.92\%) were identified as trivial. This indicated that nearly 1 in 5 packages in the NPM ecosystem could be considered trivial, highlighting their relative prevalence.

To further understand the significance and prevalence of trivial packages, their real-world usage was explored by examining download counts between June and July 2025. The analysis revealed that 12.3\% of trivial packages had over 1,000,000 downloads per month, and 28.2\% had more than 1,000 downloads per month. These statistics suggest that many trivial packages are widely used, emphasizing their importance despite their simplicity. Beyond download counts, the npms.io popularity score was also considered, which accounts for community signals like stars and forks, making it a more reliable indicator of real-world adoption~\cite{simplifyNpm}. The popularity score assigned by npms.io paints a more conservative picture. Only 6.3\% of trivial packages scored above 0.6, and a mere 0.9\% scored above 0.8 in popularity.

Taken together, these findings indicate that trivial packages are not only prevalent in the NPM ecosystem but also widely adopted, despite potentially being underrated by scoring systems that rely heavily on community interaction metrics.

\subsection{RQ2: Are trivial packages more likely to be vulnerable?} 

Previous studies have shown that trivial packages often introduce long dependency chains, which many developers cite as a major drawback of using them~\cite{Abdalkareem}. Such dependency trees increase vulnerability exposure\cite{VulandDep}, but prior work has not examined trivial packages directly.. Motivated by these observations, we conducted an empirical study to examine whether trivial packages are more likely to be vulnerable by looking at vulnerability reports via \texttt{npm audit}.

\begin{figure}[h]
\centering
\begin{subfigure}{0.29\linewidth}
    \centering
    \includegraphics[width=\linewidth]{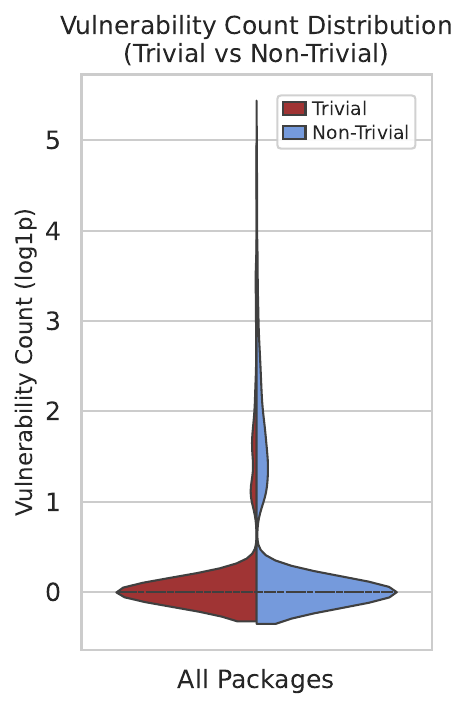}
    \caption{}
    \label{fig:vuln_comparison}
\end{subfigure}
\begin{subfigure}{0.69\linewidth}
    \centering
    \includegraphics[width=\linewidth]{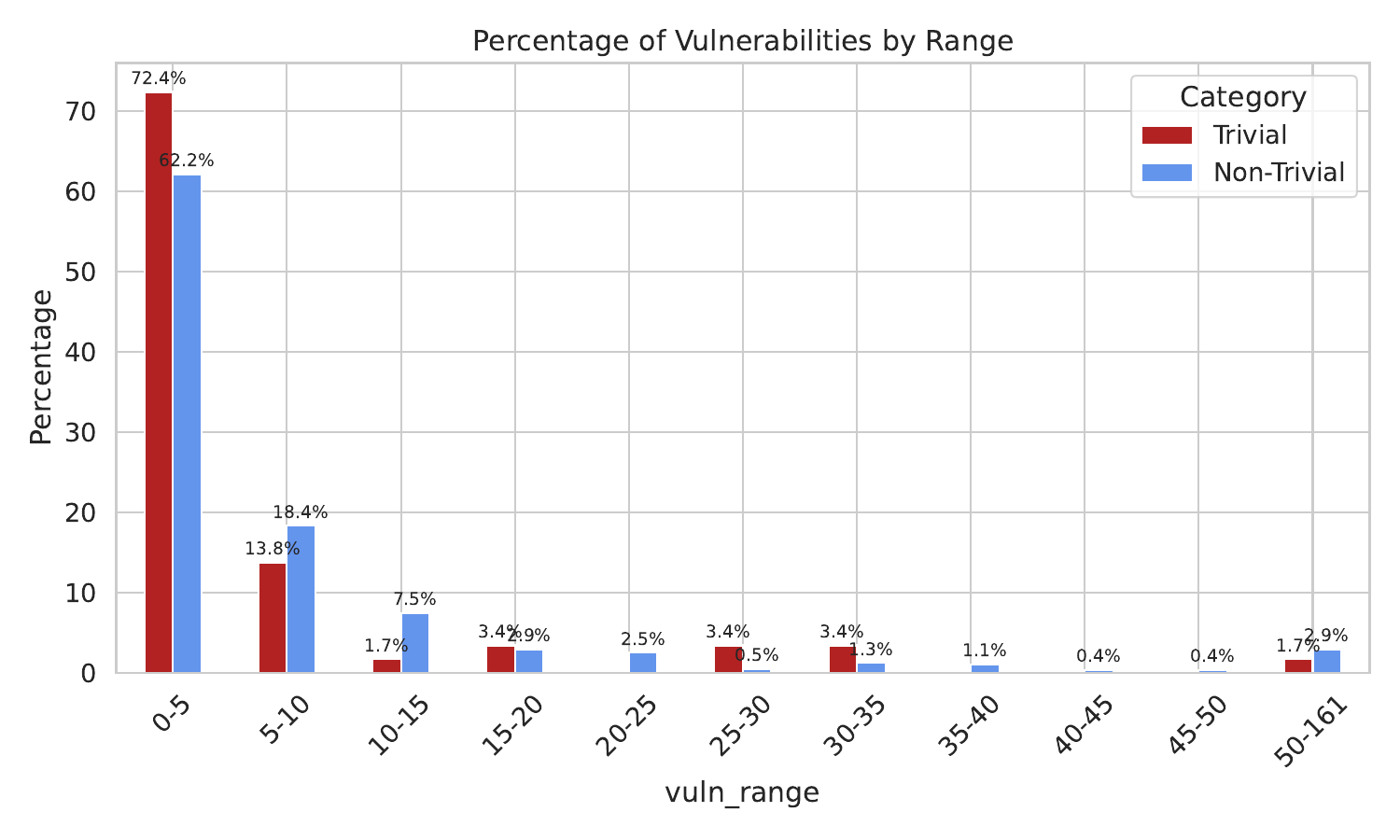}
    \caption{}
    \label{fig:vuln_percentage}
\end{subfigure}
\vspace{-0.5\baselineskip}
\caption{Trivial vs. non-trivial vulnerability distribution (a) and percentage (b).}
\end{figure}

Using the dataset of 3,220 npm packages, the relationship between package type (trivial vs. non-trivial) and known vulnerabilities was analyzed. The distribution plots shown in \autoref{fig:vuln_percentage} reveal that although most packages in both categories have fewer than five vulnerabilities, some trivial packages are outliers, with vulnerability counts exceeding 100. Overall, both package types exhibit very similar distributions, as illustrated in \autoref{fig:vuln_comparison}.

To statistically assess the difference, we performed a Mann–Whitney U test, which yielded a p-value of 0.0000000013—indicating a statistically significant difference between the two groups ($p < 0.05$). However, when Cliff’s Delta was calculated to evaluate the effect size, the result was $-0.1091$, which indicates a negligible difference. Although Non-trivial packages have slightly more vulnerabilities on average, but the difference is negligible. Trivial packages can pose similar security risks, so their inclusion may be unjustified given their potential impact on dependency chains and ecosystem stability.

\subsection{RQ3: How can we identify data-only packages automatically?}

To support further analysis, we propose a rule-based heuristic grounded in static code analysis to automatically identify both trivial and data-only packages, aligning with the study’s goal of scalable detection and risk assessment in the npm ecosystem. The assumption is that data-only packages contain no user-defined functions, based on their logic-free nature. Starting with 80 packages having zero functions, manual inspection categorized them as 40 data-only, 27 trivial, and 13 normal packages.

However, the function count alone proved insufficient. Several edge cases were encountered, including facade packages that re-export from other libraries and packages that use external dependencies without defining functions. To address this, cyclomatic complexity (normalized by file count) and dependency usage patterns were examined. Instead of relying on declared metadata, actual import behavior was analyzed by parsing source files for import and require statements, excluding internal references. Packages importing external libraries were assumed to use external logic and excluded from data-only classification. This filtering effectively removed all trivial and normal packages while retaining verified data-only packages.
\textbf{Final Rule:} Data-only packages must satisfy: \textbf{(1) function count = 0, (2) average cyclomatic complexity per file ~$\leq$~ 1, (3) no import/require statements referring to external modules.} This rule-based method enables automated detection without requiring runtime analysis or full AST parsing.

\subsection{RQ4: What are the types of data-only libraries?}\vspace{-0.5\baselineskip}

Following automated identification, the 40 data-only packages (1.24\% of the dataset) were manually classified by a single author using the zero-function, no-logic definition. Given the clarity of this definition, the process was straightforward. To understand their nature and roles, we observed two main categories:


\noindent
\textbf{Static JSON data exporters:} 
Packages exporting predefined JavaScript objects via module.exports, including large datasets and configuration modules (e.g., brittanica-r, color-name,).

\begin{figure}[h]
\centering
\begin{lstlisting}[style=mypackagecode, caption={Static JSON data exporter example from \texttt{color-name}}, label={lst:json}, captionpos=b]
    module.exports = {
        aliceblue: [240, 248, 255],
        ...
    }
\end{lstlisting}
\end{figure}

\noindent
\textbf{Constant value containers:} 
Packages exporting reusable constants like numbers or strings, designed for direct use in application logic (e.g., const-e, const-log2e).

\begin{figure}[h]
\centering
\begin{lstlisting}[style=mypackagecode, caption={Excerpt from the \texttt{const-e} package exporting only a constant value }, label={lst:const-e}, captionpos=b]
module.exports = 2.718281828459045235360287471352662497757247093699959574966;
\end{lstlisting}
\end{figure}

Despite differences in size and format, all data-only packages share common characteristics: they contain no functions, do not perform any computation or side effects, and exclusively provide static values for consumption by other packages.
    
\subsection{RQ5: Are data-only libraries more likely to be vulnerable?}

To complement the investigation of risks associated with trivial packages, this study analyzes the vulnerability exposure of data-only packages. Vulnerability data are extracted using the npm audit API, and the results are compared between data-only and non-data-only packages.

\vspace{-1\baselineskip}
\begin{table}[h]
\centering
\setlength{\tabcolsep}{10pt}
\begin{tabular}{l|c|c|c|c}
    \toprule
    \textbf{npm Packages}
    & \textbf{Min.} & \textbf{Median} & \textbf{Mean} & \textbf{Max} \\ 
    \midrule
    Data-only & 0.00 & 0.00 & 1.10 & 9.00 \\ 
    Non-Data-only  & 0.00 & 0.00 & 1.76 & 161.00 \\
    \bottomrule
\end{tabular}
\vspace{1em}
\caption{Vulnerabilities in Data-only and Non-Data-only npm Packages}
\label{tab:data_vulnerabilities}
\end{table}
\vspace{-1\baselineskip}

\vspace{-1.7em}
As shown in Table~\ref{tab:data_vulnerabilities}, data-only packages exhibit fewer vulnerabilities in raw numbers and mean values (max = 9 vs. 161, mean = 1.10 vs. 1.76), although the median remains zero for both groups. However, while the raw numbers suggest that data-only packages tend to have fewer reported vulnerabilities, statistical tests reveal no significant difference. The Mann–Whitney U test yields a non-significant result, and the effect size measured by Cliff’s Delta is negligible (0.0164),

To further confirm this result, given the large sample size imbalance (40 data-only packages vs. 3,180 non-data-only packages), bootstrapping was applied \cite{boostraping}. In 1,000 iterations, only 1.1\% of the samples showed a statistically significant difference ($p < 0.05$). This reinforces the conclusion that the difference is not statistically significant. Data-only packages, though logic-free, still carry risks from configuration or metadata, potentially exposing the ecosystem to vulnerabilities.

\subsection{RQ6: How effective is a tool to detect trivial and data-only packages?}

We implemented a command-line, rule-based analysis tool\footnote{\url{https://github.com/tnnpp/Trivial-package-detection-tool}} designed to automatically detect trivial and data-only packages in npm projects based on our defined criteria. (LOC~$\leq$~35 and complexity~$\leq$~10) and data-only packages (function count = 0, average complexity per file ~$\leq$~1, no external imports).

\vspace{-1\baselineskip}
\begin{figure}[h]
\centering
\includegraphics[width=0.45\linewidth]{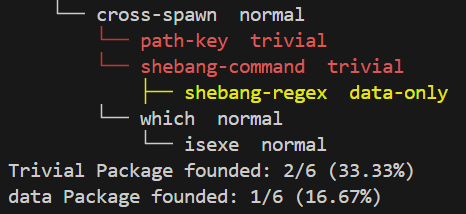}
\caption{Example tool UI}
\label{fig:tool}
\end{figure}
\vspace{-1\baselineskip}

The tool analyzes packages by calculating LOC, cyclomatic complexity, function count, and dependency usage, then classifies each as Normal, Trivial, or Data-only with a dependency tree highlighted output and summary statistics.

For evaluation, 250 samples were randomly selected using Cochran’s formula (90\% confidence). A single researcher labeled them, with ambiguous cases cross-checked by two LLMs (GPT-4o, Gemini 2.5 Flash) for additional validation and to reduce bias. The tool achieved 94\% accuracy and 0.93 weighted F1. It performed strongly for normal and trivial packages (precision 0.93–1.00, recall 0.76–0.99) but had lower recall for data-only (0.67) due to the small sample size. Overall, the rule-based tool is effective for large-scale analysis, though data-only detection needs refinement with larger datasets.

%% file: Sec5.tex
\section{Threats to Validity}

\textbf{Construct validity:} The tool has certain limitations. First, typhonjs-escomplex cannot detect some newer JavaScript syntax. Second, it does not measure lines of code or complexity in other programming languages, which may cause some packages to be misclassified. In RQ3, we evaluated the tool only on packages it could analyze. Moreover, the definition of data-only packages used in this study is a rule-based heuristic; we did not perform deeper dynamic or runtime behavior analysis. The small number of data-only cases in the dataset also limits confidence in the results for this category.
\textbf{Internal validity:} The accuracy of ground truth labeling may affect results. Although we reduced bias by consulting two LLMs for ambiguous cases, misclassifications may still occur.
\textbf{External validity:} The dataset was randomly sampled from the npm registry, but it represents only a small portion of the ecosystem and may not cover all package types.

%% file: Sec6.tex
\section{Conclusion}
Trivial and data-only packages pose overlooked risks in the npm ecosystem. In this study, our analysis found that 18\% of packages are trivial and 1.24\% are data-only, yet both can expose projects to vulnerability levels comparable to larger packages. This illustrates how package usage accelerates development time but increases maintenance and security risks. The proposed rule-based detection tool achieved 94\% accuracy, showing the feasibility of scalable static analysis. Future work will enhance detection with AST and runtime analysis, expand dataset coverage, and compare across ecosystems such as PyPI and Maven to better understand supply chain risks.